\documentstyle[12pt]{report}

\thicklines                         % thick straight lines for pictures (LaTeX)
\def\a{\alpha}
\def\b{\beta}
\def\c{\chi}
\def\d{\delta}
\def\g{\gamma}
\def\h{\eta}

\def\j{\psi}
\def\k{\kappa}
\def\l{\lambda}
\def\m{\mu}

\def\p{\pi}                     % Also, \varpi
\def\s{\sigma}                  %       \varsigma

\def\x{\xi}

\def\D{\Delta}

\def\L{\Lambda}

\def\cg{{\cal G}}
\def\ch{{\cal H}}   % overridden by cosh !!

\def\cbo{{\,\raise-.15ex\Sc [\,}}                       % curly "
\def\sl#1{\rlap{\hbox{$\mskip 1 mu /$}}#1}      % good slash for lower case
\def\Bar#1{\overline{#1}}                       % big bar
\def\lvec#1{\raisebox{0.0ex}{$\stackrel{\leftarrow}{#1}$} }  % <--  accent
\def\ddt#1{{\buildrel {\hbox{\LARGE .\kern-2pt.}} \over {#1}}}% double dot-over
\def\beqn#1{ \renewcommand{\theequation}{#1}
             \begin{equation} }
\def\eeqn{\end{equation}}
\def\beqr#1{ \renewcommand{\theequation}{#1}
             \begin{eqnarray} }
\def\eeqr{\end{eqnarray}}
\def\NON{\nonumber\\}
\def\beqrabc#1{ \setcounter{equation}{0}
                \renewcommand{\theequation}{#1\alph{equation}}
                \begin{eqnarray} }
\def\beqrn#1#2{ \setcounter{equation}{#2}
                \renewcommand{\theequation}{#1.\arabic{equation}}
                \begin{eqnarray} }

\def\frac#1#2{ {\sstyle {#1\over #2} } }
\def\det#1{{\rm det}\left(#1\right)}

\def\NPB#1{Nucl. Phys. {\bf B#1}}
\def\PLB#1{Phys. Lett. {\bf B#1}}

\def\PRL#1{Phys. Rev. Lett. {\bf #1}}

\def\sstyle{\scriptstyle}

\def\rhs{\mbox{r.h.s.} }

\def\ie{\mbox{i.e.} }
\def\eg{\mbox{e.g.} }

\def\half{{1\over 2}}
\def\Re{{\rm Re}}
\def\Im{{\rm Im}}
\def\ch{{\rm cosh}}
\def\sh{{\rm sinh}}
\def\tgh{{\rm tgh}}
\def\det{{\rm det}}
\def\Det{{\rm Det}}

\def\hdf{\hat{D}_F}

\begin{document}
\noindent December 1992 \hfill WIS--92/97/12--PH
\par
\begin{center}
\vspace{15mm}
{\large\bf The Euclidean Spectrum of\\
Kaplan's Lattice Chiral Fermions}\\[5mm]
{\it by}\\[5mm]
Yigal Shamir\\
Department of Physics\\
Weizmann Institute of Science, Rehovot 76100, ISRAEL\\[15mm]
{ABSTRACT}\\[2mm]
  \end{center}
\begin{quotation}
 We consider the $(2n\! +\! 1)$-dimensional euclidean Dirac operator
with a mass term that looks like a domain wall,
recently proposed by Kaplan to describe chiral
fermions in $2n$ dimensions. In the continuum case we show that the
euclidean spectrum contains {\it no} bound states with non-zero momentum.
On the lattice, a bound state spectrum without energy gap exists
only if $m$ is fine tuned to some special values, and the
dispersion relation does not describe a relativistic fermion.
In spite of these peculiarities, the fermionic propagator
{\it has} the expected $1/\sl{p}$ pole on the domain wall.
But there may be a problem with the phase of the fermionic determinant
at the non-perturbative level.
\end{quotation}

\vspace{30mm}
\noindent email: ftshamir@weizmann.bitnet
\newpage

  Numerous attempts~[1] in the past to define chiral gauge theories using
lattice regularization have been unsuccessful because of the doubling
problem~[2]. Recently, Kaplan~[3] proposed that a lattice chiral gauge theory
in
$2n$ dimensions could arise as the low energy limit of a massive Dirac theory
in $2n+1$ dimensions, provided the fermion mass takes the form of
a domain wall
$$ m(s)=m\,sign(s) \,,\eqno(1)$$
where $s$ denotes the extra coordinate. Besides massive
$(2n\! +\! 1)$-dimensional excitations, the hamiltonian has a spectrum of bound
states~[3-5]
$$ \j(x,s;p)=e^{-m|s|+ip\cdot x}u(p) \,. \eqno(2)$$
Here $x_\m$ denote the $2n$-dimensional coordinates and $u(p)$ is a
positive chirality spinor satisfying $\sl{p}u(p)=0$.
Eq.~(2) describes a massless chiral fermion that lives
on the domain wall.

  A naive discretization of the continuum hamiltonian does not give rise to
chiral fermions because the lattice zero mode has the expected right handed
doubler. However, Kaplan~[3] and Jansen and Schmaltz~[6] showed that by adding
a Wilson term one can obtain a chiral spectrum for a certain range of the
ratio $m/r$ where $r$ is the Wilson parameter. For
$0<|m/r|<2$ the free lattice hamiltonian describes one Weyl fermion near the
origin of the Brillouin zone. For higher $|m/r|$ the chiral fermions move
to the vicinity of other corners of the Brillouin zone.

  Thanks to current conservation which is always true in odd dimensions, the
$2n$-dimensional anomaly can be written as
$\partial_\m j_\m = -\partial_{2n+1} j_{2n+1}$. Thus, the anomaly arises as
a non-zero flux away from the domain wall.
The flux has been calculated~[7] for general $m/r$ in the presence of an
external, slowly varying gauge field. It is a non-trivial result that the
flux takes only integral values, and it changes discontinuously as a function
of $m/r$ in agreement with the anomaly anticipated from the lattice chiral
spectrum.

  In this paper we consider the Dirac operator of the euclidean theory.
Understanding the properties of the Dirac operator is important for obvious
reasons. The euclidean partition function is better defined mathematically
than the Minkowskian path integral, and
numerical simulations rely heavily on the euclidean metric for convergence.

  Our discussion is limited to the Dirac operator of the free
theory, its only non-trivial feature being the domain wall shape of the fermion
mass. We  first discuss the properties of the {\it continuum}
Dirac operator. Surprisingly, we find that its spectrum
is drastically deferent from the spectrum of the hamiltonian.
The most important result is that there exist {\it no} bound states with
non-zero momentum! Another peculiarity is that the zero momentum
spectrum contains an {\it infinite}
number of bound states. For general momentum, the spectrum consists of
eigenstates which are superposition of plane waves in one half of the
$(2n\! +\! 1)$-dimensional space and which decay exponentially in
the other half. Only these eigenstates
contribute to physical observables because $p_\m=0$ is
an isolated discontinuity point of the spectrum.

  The peculiar behaviour described above is possible because the euclidean
Dirac operator is not hermitian. For hermitian operators such as the
hamiltonian, the existence of a zero momentum bound state separated by an
energy gap from the massive modes implies that finite momentum bound states
must also exist. However, the variational argument needed to prove the above
statement fails when the eigenvalues are complex.

  On the lattice we discuss in detail the case $r=1$. Taking a finite lattice
with periodic boundary condition in the $s$ direction we find yet another
surprise. For general $m$, even the $p_\m=0$ spectrum contains
no (approximate) zero mode.
Most of the eigenstates {\it do} look like bound states, but
the eigenvalues are typically all $O(1)$.
The only exception is when $m$ takes the special value $m=\sqrt{2}$.
In this case one has a bound state spectrum with no energy gap,
but the small $p$ dispersion
relation is quadratic rather than linear. (The exact form of the dispersion
relation is non-local). The case $r\ne 1$ is technically more complicated
and we do not discuss it in detail.
We expect that the qualitative properties of the spectrum should be the
same as in the $r=1$ case.

  The eigenstates of the euclidean Dirac operator do not have simple
physical interpretation. What enters the calculation of physical
correlation functions is the fermionic propagator and the fermionic
determinant. The continuum fermionic propagator can be constructed
by considering
the operator $D_F D_F^\dagger$. Denoting by $\cg$ the propagator of
$D_F D_F^\dagger$, the fermionic propagator $\cg_F$ is
given by $\cg_F=D_F^\dagger\,\cg$.

  For $p_\m=0$, $D_F D_F^\dagger$ has a {\it negative} chirality zero mode.
Since $D_F D_F^\dagger$ is hermitian, the zero momentum
bound state implies the existence of a bound state spectrum for $p_\m\ne 0$
as well, which, in turn,
gives rise to a $1/p^2$ pole in $\cg$ on the domain wall.
Consequently, $\cg_F$ {\it has} the desired chiral $1/\sl{p}$
pole on the domain wall in spite of the absence of finite
momentum bound states in the spectrum of $D_F$!

  On a finite lattice, $D_F$ is a matrix and $D_F D_F^\dagger$ is defined
through matrix multiplication.
Because of the Wilson term, the lattice $D_F D_F^\dagger$
may contain up to fourth lattice derivatives, and so
arguments based on the operator $D_F D_F^\dagger$ are more delicate.
In the case $r=1$ this difficulty does not arise. The lattice
$D_F D_F^\dagger$ contains only second derivatives, and its only unusual
feature is that the normalization of the kinetic term is different in the
regions $s>0$ and $s<0$. Thus, the argument goes through and,
at least in the $r=1$ case, we can prove that the lattice propagator has a
chiral $1/\sl{p}$ pole on the domain wall when $m$ is in the allowed range~[6].

  A similar argument applies to the
modulus of the fermionic determinant. It, too, can be constructed from
$D_F D_F^\dagger$ because $|\Det\{D_F\}|=\Det^\half\{D_F D_F^\dagger\}$.
When gauge fields are introduced, one can expect
that the response of fermionic propagator and
the modulus of the fermionic determinant will be dominated
by the chiral mode of $D_F D_F^\dagger$.

  The object which must be obtained directly from $D_F$ is the {\it phase}
of the fermionic determinant. The peculiar spectrum of $D_F$ may indicate
an unexpected behaviour of the phase of $\Det\{D_F\}$.
Naively, one can imagine several scenarios
according to which the unusual properties of the euclidean spectrum may
prevent the existence of a chiral continuum limit. Finding what really
happens requires a detailed understanding of the response of the phase of
$\Det\{D_F\}$ to gauge fields.
We comment that the Chern-Simons action found in ref.~[7],
which reproduces the usual anomaly cancellation condition, arise in weak
coupling perturbation theory. However, we expect that
the continuum limit should depend on the {\it dynamical} properties of the five
dimensional gauge theory, and hence that non-perturbative
contributions should play an important role.

\vspace{3ex}

  We begin our analysis with the continuum Dirac operator
$$ D_F=i\g_\m\partial_\m + i\g_5\partial_s + im(s) \,.\eqno(3)$$
For simplicity we work in $(4+1)$ dimensions. All gamma matrices are
hermitian and repeated indices are summed from one to four. Most of
the discussion generalizes trivially to other dimensions. The dependence
on dimensionality will be explicitly given whenever necessary.

  Assuming a plane wave solution in the first four coordinates we obtain
an effective one dimensional problem
$$ \hdf=i\gamma_5\partial_5 + im(s) + \sl{p} \,.\eqno(4)$$
We consider the eigenvalue equation
$$ \hdf\j=\l\j \,.\eqno(5)$$
The general solution can be written as
$$ \j=U(s;\l,p)u(\l,p,\a) \,,\eqno(6)$$
where $U(s;\l,p)$ is a four by four matrix and $u(\l,p,\a)$ is an
$s$-independent spinor. $\a$ is a spin index which we will usually suppress.
Integrating eq.~(5) we find
$$ U(s)=\left\{
\begin{array}{cc}
\ch(s\k_+) + \k_+^{-1}\g_5(i(\sl{p}-\l)-m)\sh(s\k_+)\,, & s\ge 0\,, \\
\ch(s\k_-) + \k_-^{-1}\g_5(i(\sl{p}-\l)+m)\sh(s\k_-)\,, & s\le 0\,,
\end{array}\right.
\eqno(7)$$
where
$$ \k_\pm^2=(i\l\pm m)^2 + p^2 \,.\eqno(8)$$
For definiteness, we pick the branch were
$\k_\pm\to i\l\pm m$ for $p^2 \to 0$. Notice that $U(s)$ is continuous
at $s=0$ and has a discontinuous derivative there as required by eq.~(4).
One can obtain an analytic dependence on $s$ by smoothing the
discontinuity of the mass function at $s=0$. However, the qualitative
properties of the euclidean spectrum depend only on the asymptotic
behaviour of $m(s)$ at $\pm\infty$, and they are completely insensitive to
the details of the transition region.

  For general values of $\l$, the solution $U(s)$ contains exponentially
increasing and exponentially decreasing pieces.
We begin our investigation
by looking for special values of $\l$ for which the solution is a
superposition of plane waves in some region of the $s$-axis.
Demanding for example $\Re\k_+=0$ we find that $\l$ must take the special
form $\l=\x+im$ where the real parameter $\x$ satisfies $\x^2\ge p^2$.
In the region $s>0$, $U(s)$ is then a superposition of plane waves
with $k=\pm(\x^2-p^2)^\half$. To obtain $U(s)$ for
$s<0$ we need $\k_-$. Assuming $\x^2,p^2\ll m^2$ we find
$\k_-\approx -2m-p^2/4m+i\x$. Hence, the solution has no plane
wave component in the region $s<0$. For given $p_\m$ and $\l$ there are
four linearly independent solutions, of which two increase exponentially
and two decrease exponentially for $s<0$. These solutions are
obtained by an appropriate choice of the constant spinor $u(\l,p)$, see
below.

  The spectrum is obtained by first taking
$s$ to lie in the finite interval $-L\le s \le L$ with some local
boundary conditions at $s=\pm L$. We obtain the spectrum below assuming
periodic boundary conditions. (Other choices of the
boundary conditions lead to similar conclusions). In the limit $L\to\infty$,
only solutions which decrease exponentially for $s<0$ remain
in the spectrum. Of course, there is another family of solutions for
which the role of $s>0$ and $s<0$ is interchanged.

  Let us now examine the solution of eq.~(5) for $|\Im\l|<m$.
In this case one has $\Re\k_+>0$
and $\Re\k_-<0$. If no special choice of $u(\l,p)$
is made, the solution will contain exponentially increasing and exponentially
decreasing pieces for both $s>0$ and $s<0$.
To obtain only exponentially
decreasing (respectively increasing) solution for $s>0$ we demand
$$ \L^>_\pm u(\l,p)=0 \,,\eqno(9)$$
where
$$ \L^>_\pm=\half\pm\g_5 {i(\sl{p}-\l)-m\over 2\k_+} \,.\eqno(10)$$
Notice that $\L^>_\pm$ are orthogonal projection operators.
Similarly, taking into account that $\Re\k_-<0$, an exponentially
decreasing (respectively increasing) solution for $s<0$
is obtained by demanding
$$ \L^<_\pm u(\l,p)=0 \,,\eqno(11)$$
where
$$ \L^<_\pm=\half\pm\g_5 {i(\sl{p}-\l)+m\over 2\k_-} \,.\eqno(12)$$

  A normalizable eigenstate is obtained by demading that the solution
decrease exponentially on both sides of the domain wall, \ie
$u(\l,p)$ must satisfy $\L^>_+ u(\l,p)=\L^<_+ u(\l,p)=0$ simultaneously.
Using eqs.~(10) and~(12) we find that a non-trivial solution exists if and
only if the matrix $(\k_+ -\k_- -2m\g_5)$ has a zero eigenvalue.
For this to happen,
we must have $\k_+ -\k_- =2m$, which, in turn, is true only when $p_\m=0$.
In this case $(\k_+ -\k_- -2m\g_5)$ becomes a projection operator and the
bound state has a definite chirality.

  We therefore arrive at the conclusion that euclidean bound states
exist only for $p_\m=0$. When $p_\m\ne 0$, one can obtain solutions
that decrease exponentially on one side of the domain wall but not on both.
The technical reason for this behaviour is that, as can be seen from eq.~(4),
in euclidean space $\sl{p}$ couples the positive and negative chiralities.
(By contrast, the kinetic term of the hamiltonian $\vec\a\cdot\vec{p}$
{\it commutes} with $\g_5$). The homogeneous solution with negative chirality
increase exponentially on both sides of the domain wall, and for finite $p_\m$
one can eliminate the exponentially growing component only on one side.

  For $p_\m=0$ the positive and negative chiralities decouple, and
$U(s)$ simplifies to
$$ U(s)=\half(1+\g_5)e^{-m|s|-i\l s} + \half(1-\g_5)e^{m|s|+i\l s}\,
\eqno(13)$$
We readily see that for $p_\m=0$ there is a positive chirality bound state
for {\it all} $\l$ provided $|\Im\l|<m$.

  Our next step is to obtain the spectrum by imposing periodic boundary
conditions in the finite interval $-L\le s \le L$, later taking the limit
$L\to \infty$.
Imposing periodic boundary condition amounts to introducing
an anti-domain wall at $s=L$. By the same argument as before,
bound states which live on the anti-domain wall exist only for $p_\m=0$.
These bound states correspond to the negative chirality part of eq.~(13).

  Eigenstates which satisfy periodic boundary conditions occur when the
matrix equation
$$ \left( U(s=L)-U(s=-L) \right) v=0 \,,\eqno(14)$$
has a non-trivial solution. The zero eigenvectors are than identified with
the spinors $u(\l,p,\a)$. We must therefore look for the zeros of
$\det \left( U(L)-U(-L) \right)$. After some algebra we find in $2n+1$
dimensions
$$ \det \left( U(L)-U(-L) \right)=\D^{2^{n-1}} \,,\eqno(15)$$
where
$$ \half\D= 1 - \ch(L\k_+)\ch(-L\k_-) +
   {p^2-m^2-\l^2 \over \k_+\k_-} \sh(L\k_+)\sh(-L\k_-)
\,.\eqno(16)$$

  Consider first the special case $p_\m=0$. In this case further simplification
occurs and we obtain
$$ \half\D= 1 - \ch(2i\l L) \,.\eqno(17)$$
The zeros occur for $\l_k=\p k/L$. Thus, imposing the periodic boundary
conditions eliminates all bound states with $\Im\l\ne 0$ from the spectrum.
However, we are still left with an infinite number of bound states, which
becomes a continuum of bound states where $\l$ is any real number in the
limit $L\to\infty$.

  The existence of infinitely many bound states for $p_\m=0$ provides another
explanation for the absence of finite momentum bound states. Had every $p_\m=0$
bound state been the end point of a continuous four dimensional spectrum,
we would have found that the number of fields that live on the (anti)-domain
wall is infinite! We comment that, the fact that the
eigenvalues lie on the real axis is a consequence of the symmetry between
the two sides of the domain wall. Had $|m(s>0)|$ been different
from $|m(s<0)|$, we would have found that the bound states' eigenvalues move
away from the real axis. We will see that a similar phenomenon takes place
on the lattice.

  We next consider $\D$ for $p_\m\ne 0$. We will show that the zeros of
$\D$ occur for $|\Im\l|\approx m+O(1/L)$. Let us first assume that
$|\Im\l|-m$ is large in units of $1/L$. In this case $\D$ is dominated by
the exponentially increasing part of the $\ch$ and $\sh$ functions.
Therefore, a necessary condition for having $\D=0$ is that the
difference between the coefficient of the $\ch$ term and the coefficient
of the $\sh$ term be exponentially small. Explicitly, we must have
$\k_+^2\k_-^2=(p^2-m^2-\l^2)^2+O(e^{-mL})$.  This condition is satisfied only
if $p/m\approx e^{-mL}$.

  In the hamiltonian picture, boundary effects on the chiral
mode are $O(e^{-mL})$. Thus, for exponentially small momentum we expect
a significant distortion of the chiral mode and so $p/m\approx e^{-mL}$
is of no physical interest.

  Assuming $p/m\gg e^{-mL}$, we conclude that $\D$ has no zeros if
$|\Im\l|-m$ is large.
The zeros of $\D$ must therefore be found in the vicinity of $|\Im\l|=m$.
For definiteness let us assume $\Im\l\approx m$. A convenient parametrization
is $\l=\x+i(m-k\b/\x L)$. (We take $k=+(\x^2-p^2)^\half$ and $\x^2\ge p^2$
as before). This implies $\k_+\approx ik+\b/L$. Now, $\ch(L\k_+)$
and $\sh(L\k_+)$ are $O(1)$, whereas
$\ch(-L\k_-)\approx\sh(-L\k_-)\approx\exp(-L\k_-)$. (Notice that
$\Re\k_-$ is negative). Hence
$$ \D\approx \exp(-L\k_-) \left(\ch(L\k_+) +
   {p^2-m^2-\l^2 \over \k_+\k_-} \sh(L\k_+) \right) \,.
\eqno(18)$$
Up to exponentially small corrections, the zeros of $\D$ occur for
$$ \tgh(L\k_+)={ \k_+\k_- \over p^2-m^2-\l^2 }\,.\eqno(19)$$
We can find the zeros explicitly in the limit $p^2,\x^2\ll m^2$. We first
notice that
$$ \tgh(L\k_+)={\sh(2\b)+i\sin(2Lk) \over \ch(2\b)+\cos(2Lk)} \,,\eqno(20)$$
whereas
$$ { \k_+\k_- \over p^2-m^2-\l^2 }={k\over\x}+O(1/m) \,.\eqno(21)$$
Since $k/\x$ is real, the zeros occur when $\sin(2kL)=0$.  We than
determine $\b$ from the equation
$$ {\sh(2\b) \over \ch(2\b)\pm 1}={k\over\x} \,.\eqno(22)$$

  The finite $L$ eigenstates have the following shape. The spinor $u(\l,p,\a)$
is made almost entirely out of the $\L^<_+$ projection, plus an
exponentially small contribution from the $\L^<_-$ projection.
When we go from $s=0$ into the negative $s$ region, the wave
function first decreases exponentially. Around $s=-L/2$ the other solution
takes over, and the wave function starts growing exponentially, until at $s=-L$
it becomes $O(1)$ again. In the limit $L\to\infty$, $(\Im\l-m)$ vanishes, the
exponentially growing region recedes to $-\infty$ and we obtain the infinite
space spectrum described earlier.

\vspace{3ex}

  We now turn to the lattice Dirac operator.
We take a lattice of $2N$ sites in the $s$ direction with periodic
boundary conditions, \ie the points $s=N$ and $s=-N$ are identified.
We set $m(s)=m$ for $s=1,\ldots,N$ and $m(s)=-m$ for $s=-N+1,\ldots,0$.
(Alternatively, we could take $m(s)=0$ at the sites $s=0$ and $s=N$. For
large $N$ this modification has a negligibly small effect).

  As in the continuum case we assume a plane wave solution in the
first $2n$ coordinates. The effective Dirac operator in the $s$
direction becomes an $8N\times 8N$ matrix.
With lattice spacing $a=1$, the eigenvalue equation reads
\beqr{23}
   \half\g_5\left(U(s+1)-U(s-1)\right)
   +{r\over 2}\left(U(s+1)+U(s-1)-2U(s)\right) & & \NON
   +\left(i\sum_{\m=1}^{2n}\g_\m \sin(p_\m)-rF(p)+m(s)+i\l\right)U(s)
   & = & 0 \,,
\eeqr
$$ F(p)= \sum_{\m=1}^{2n} (1-\cos(p_\m)) \,.\eqno(24)$$
Here $U(s)$ is a four component spinor. As before $\l$ is the euclidean
eigenvalue.

  We consider first the case $r=1$. Denoting by $U_\pm(s)$ the positive
(negative) chirality components, Eq.~(23) simplifies to
\beqrabc{25}
  U_+(s+1) & = & \left( 1+F(p)-i\l-m(s) \right)U_+(s)
                 -i \sum_{\m=1}^{2n}\s_\m \sin(p_\m) \, U_-(s) \,, \\
  U_-(s-1) & = & \left( 1+F(p)-i\l-m(s) \right)U_-(s)
                 -i \sum_{\m=1}^{2n}\Bar\s_\m \sin(p_\m) \, U_+(s) \,.
\eeqr
  We introduce transfer matrices defined by
\beqrabc{26}
  U(s+1) & = & T^+\, U(s)\,, \quad\quad s>0 \,, \\
  U(s-1) & = & T^-\, U(s)\,, \quad\quad s\le 0 \,.
\eeqr
To obtain \eg $T^+$ we use eq.~(25b) to express $U_-(s+1)$ in terms of
$U_-(s)$ and $U_+(s+1)$, and eq.~(25a) to express $U_+(s+1)$ in terms of
$U_+(s)$ and $U_-(s)$. We find
$$ T^+ = \left( \begin{array}{cc}
      1+F(p)-i\l-m &
      -i \raisebox{.0ex}[3ex][3ex]{$\sum_{\m=1}^{2n}$} \s_\m \sin(p_\m)    \\
      i \raisebox{.0ex}[3ex][3ex]{$\sum_{\m=1}^{2n}$} \Bar\s_\m \sin(p_\m) &
      {1+G(p)\over 1+F(p)-i\l-m}
   \end{array}\right)\,,
\eqno(27)$$
where
$$ G(p)=\sum_{\m=1}^{2n} \sin^2(p_\m) \,.\eqno(28)$$
Notice that $\det(T^+)=1$. To obtain $T^-$, invert $T^+$ and replace $m$ by
$-m$.

  We will analyze the spectrum in the vicinity of the origin of the Brillouin
zone. We assume $0<m<2$ so that on an infinite lattice the zero mode~[3] indeed
occurs at $p_\m=0$. For larger values of $m$ the zero modes~[6] occur at
other corners of the Brillouin zone, but the analysis
remains essentially the same.

  We begin with the $p_\m=0$ case. The positive and negative chiralities
decouple and we consider the positive chirality eigenvectors.
We will show that the
positive chirality zero mode~[3] of the infinite lattice disappears on a
finite lattice (no matter how large) unless we fine tune $m$ to a special
value. In the positive chirality channel, the periodicity requirement reads
$$ (1-i\l-m)^N=(1-i\l+m)^{-N} \,.\eqno(29)$$
The eigenvalues which solve eq.~(29) are
$$ i\l_k^\pm=1\pm \left( m^2+e^{{2\p ik\over N}} \right)^\half
   \,,\quad\quad k=0,\ldots,N-1\,.
\eqno(30)$$
The eigenvectors are given by $U_+(s+1)=(1-i\l-m) U_+(s)$ for $s>0$ and
$U_+(s-1)=(1-i\l+m)^{-1} U_+(s)$ for $s\le 0$.
Notice that most of the eigenvectors look like bound states which live
on one of the domain walls.

  A zero mode occurs only if $N$ is even and $m=\sqrt{2}$. For arbitrary $m$,
the eigenvalues are all $O(1)$. This has the following explanation.
On an infinite lattice, a euclidean bound state exists for any $\l$ which
belongs to the intersection of the open sets defined by $|1-i\l-m|<1$ and
$|1-i\l+m|>1$. (For $m$ in the allowed range, the intersection always contains
a neighbourhood of the origin in the complex plane).
This region is the analog of the stripe $|\Im\l|<m$ in the continuum case.
The finite lattice's boundary conditions pick a finite number of points out
of this region. The Wilson term breaks the
symmetry between the two sides of the domain wall, and so, unlike the continuum
case, the eigenvalues do not have any special reality property.

  Since the lattice Dirac operator is a finite matrix, its eigenvalues and
eigenvectors solve algebraic equations and so they must depend continuously
on $p_\m$. The magnitude of the deviation
from the $p_\m=0$ behaviour is controlled by the parameter $|p_\m N|$.
For $p_\m\ll N^{-1}$ the distortion will be small,
whereas for $p_\m\ge N^{-1}$ we expect that the shape of
the eigenvectors will be qualitatively different from the $p_\m=0$ case.

  We now want to find whether the lattice supports a spectrum of bound states
for $N^{-1}\le p_\m\ll 1$.
The analysis parallels the continuum case (although the
results are different) and we only outline it here.
The transfer matrix $T^+$ has two eigenvalues which
satisfy $\h^+_1 = 1/\h^+_2$. (In $2n+1$ dimensions the multipliticity of
each eigenvalue is $2^{n-1}$).
Without loss of generality we assume that
$|\h^+_1|<1$ .  A similar statement applies to $T^-$.

  We define projection operators into the subspaces that belong to each
eigenvalue separately for $T^+$ and $T^-$.
In order to have a bound state, the $\h^+_1$
and $\h^-_1$ subspaces must have a non-trivial intersection. Calculating
to leading order in $p_\m$ and $\l$ we find the necessary condition
$$ 2-m^2+2p^2-2i\l\approx 0 \,.\eqno(31)$$
Eq.~(31) holds only if $m=\sqrt{2}$ and $i\l\approx p^2$.
In order to verify that the bound states
belong to the spectrum we must check that they satisfy periodic boundary
condition. It is easy to verify that, provided
$$ i\l=1+F(p)-\left(1-G(p)\right)^\half \,,\eqno(32)$$
one has $T^+=-T^-$. Thus, taking $N$ to be even we conclude that the bound
states belong to the spectrum and that the exact dispersion relation is given
by eq.~(32). We see that, in the special case where the lattice spectrum
has an exact zero mode, this zero mode is the end point of a
bound state spectrum.

  The bound state spectrum found above describes a lattice excitation that
lives
on the domain wall. But it is clear that this mode does not describe
a relativistic fermion. What the euclidean mode~(32) has in common with the
chiral mode of the hamiltonian is (a)  the absence of an energy gap
(b) the existence of two states per given momentum
and (c) the absence of other zeros except the one at $p_\m=0$.
But there are also important differences. The euclidean mode has a definite
chirality only at $p_\m=0$. The absence of an energy gap depends on fine
tuning the mass parameter, whereas the masslessness of the chiral mode of the
hamiltonian arise because of topological reasons. Finally, the dispersion
relation~(32) is non-local, and for small $p_\m$ it is {\it quadratic} rather
than {\it linear}.

  The mass function~(1) is of particularly simple shape, and one could consider
the effect of a more general ansatz. Since the fifth
direction is not physical, in the most general case the mass function may
depend on chirality and may contain non-local couplings in the $s$
direction. The only absolute requirement is that $m(s)$ respect locality and
spacetime symmetries in four dimensions. Remarkably, the behaviour exhibited
by the mode~(32) seems to persists in the most general case in euclidean
space.

  In trying to understand this behaviour, we notice that properties (a-c) above
can be satisfied without even introducing a fifth dimension. All That is needed
is to take the {\it four dimensional} massless Dirac operator and to add a
``chiral'' mass term $\half(1-\g_5)m$. The reader can easily check that
the small $p_\m$ dispersion relation is then quadratic instead of linear.
Obviously, in this case the no-go theorems apply, and the fact that the
small $p_\m$ dispersion relation is quadratic is crucial in order to avoid
additional zeros inside the Brillouin zone.

  In the present context we have a five dimensional
problem. However, since we are dealing with free fermions, picking a
particular mode automatically defines an effective (in general non-local)
four dimensional theory. Concentrating on the small $p_\m$ behaviour,
whenever we demand that this mode
satisfy properties (a) and (b) above, we end up with an effective
Dirac operator that has the ``chiral'' mass term, with the resulting quadratic
dispersion relation.

  We comment that the lattice {\it hamiltonian} evades the no-go
theorems because of the possibility of level crossing. As in four dimensions,
the spectrum must be periodic across the Brillouin zone, but when we
keep track of the flow of a particular eigenvalue as a function of $p_\m$, we
find ourselves at $p=2\p$ on a different branch from the one we started with.
We have already seen that euclidean bound states have little to do with the
bound states of the hamiltonian. It will be interesting to check whether
in the present context one can prove a no-go theorem only for the euclidean
Dirac operator.

  We finally comment that the case $r\ne 1$, while being technically more
complicated, readily
contains all the ingredients that make the euclidean spectrum so different
from one's naive expectation. In particular, we expect that most of the
small $p_\m$ eigenvectors will look like bound states, but that a massless mode
will occur only incidentally. Furthermore, whenever such a mode should occur,
it is bound to suffer from the same problem as discussed above, \ie to have a
quadratic instead of a linear dispersion relation.

\vspace{3ex}

  After this lengthy discussion of the peculiarities of the euclidean
spectrum, the demonstration that the fermionic propagator {\it has} the desired
$1/\sl{p}$ pole on the (anti)-domain wall is remarkably simple.
We consider first the continuum case. One has
$$ D_F D_F^\dagger=-\partial_s^2-\partial_\m\partial_\m+m^2+2m\g_5\d(s)
\,.\eqno(33)$$
In an infinite space  $D_F D_F^\dagger$ has a bound state spectrum
$$ \j(x,s;p)=e^{-m|s|+ip\cdot x} u(p) \,,\eqno(34)$$
where $u(p)$ is an arbitrary {\it negative} chirality spinor, and
$$ D_F D_F^\dagger \,\j(x,s;p)= p^2 \j(x,s;p) \,.\eqno(35)$$
The dispersion relation is therefore $\l^2=p^2$. It is strictly local,
reflecting the separation of variables which characterizes the mode~(34).
The contribution of the mode~(34) to the propagator of $D_F D_F^\dagger$ is
$$ \cg^{chiral}(s,s';p)=\half(1-\g_5){e^{-m|s|-m|s'|}\over p^2} \,.\eqno(36)$$
Since $\half(1-\g_5)e^{-m|s|}$ is annihilated by $D_F^\dagger$, the
contribution to the fermionic propagator is
\beqr{37}
  \cg^{chiral}_F(s,s';p) & = & D_F^\dagger\, \cg^{chiral}(s,s';p) \NON
  & = & \half(1+\g_5){e^{-m|s|-m|s'|}\over\sl{p}}\,.
\eeqr
Clearly, eq.~(37) describes the propagation of a chiral
fermion on the domain wall.

  When we impose periodic boundary conditions, the propagator develops another
pole which describes a Weyl fermion with the opposite chirality that lives on
the anti-domain wall. The crucial point is that, using the hermiticity of
$D_F D_F^\dagger$ we can rigorously show that the chiral poles are modified
at most by $O(e^{-mL})$ corrections. In fact, in the present case there is
no modification at all because the mode~(34) is symmetric under $s\to\,-s$.

  As we have explained in the introduction, similar conclusions apply to
the lattice propagator in the case $r=1$. On the lattice the boundary effects
are $O(e^{-N})$. They do not vanish identically because of the absence of
an $s\to\,-s$ symmetry. We have not analyzed in detail the case $r\ne 1$.
In any event, since the lattice effects tend to zero in the limit of
small $m$ and $p_\m$, we expect that
the continuum behaviour should persist for some finite range of $m$,
if not for the entire interval found in ref.~[6].

\vspace{3ex}

  We conclude with a summary of our main results. Any object which can be
obtained from the hermitian operator $D_F D_F^\dagger$ is ``well behaved''.
In particular, the fermionic propagator has the expected chiral poles at
least for some range of $m$ and $r$ values.
Since perturbation theory is defined as the sum of one loop
diagrams, we expect that weak coupling perturbation theory on the lattice
should lead to the same conclusions as found  when perturbation theory is
defined through dimensional regularization. Thus, our conclusions
support the results of ref.~[7].

  At the non-perturbative level we still expect the modulus of the fermionic
determinant to be well behaved. But the phase of
the fermionic determinant may exhibit a peculiar behaviour because it must be
constructed directly from the spectrum of $D_F$ itself.

 To make our point more clear, consider the construction of the fermionic
propagator {\it directly} from the spectrum of $D_F$. We remind the reader
that,
being complex, $D_F$ has ``right eigenstates'' and ``left eigenstates''.
So far we have discussed only the right eigenstates, which are defined by
the eigenvalue equation~(5). The left eigenstates are obtained when we take
$D_F$ to act to the left, \ie
$$ \Bar\c \,\lvec{D}_F = \l\,\Bar\c \,.\eqno(38)$$
(On the lattice, the left action of $D_F$ is represented by the
transposed matrix). The left and right eigenstates are mutually orthogonal
$$ \mbox{\bf (} \Bar\c_{i\a} | \j_{j\b} \mbox{\bf )} =
   \left\{\begin{array}{cc}
      0\,, & \l_i\ne\l_j\,, \\
      \d_{\a\b} \,, &  \l_i=\l_j\,.
\end{array}\right.
\eqno(39)$$
The indices $\a$ and $\b$ count the multipliticity of each eigenvalue.
The propagator has the generic form
$$ \cg_F = \sum_{i,\a} {\j_{i\a}(x,s)\,\Bar\c_{i,\a}(x',s')\over\l_i}
\,.\eqno(40)$$

  As we have explained earlier,  the pole in the propagator should persist
own to $p \approx e^{-mL}$. This means that in the limit of small $p_\m$, the
{\it norm} of $\cg_F$ can be very large.
In the continuum case, the propagator is constructed from the balk
eigenstates of the two half-spaces. The interesting region is
$ e^{-mL}\ll p/m \ll 1$. Since $|\Im\l|\approx m$,
the denominator on the \rhs of eq.~(40) is never small.
This observation means that, in order to build the pole singularity,
balk eigenstate up to $\Re\l\approx m^2/p$ must contribute {\it coherently}
to the fermionic propagator in the vicinity of the domain walls. Thus, the
ultraviolet and the infrared behaviour of the euclidean spectrum are
closely related.

  On a finite lattice this mechanism cannot work, because the number
of eigenstates per given momentum is just $8N$, whereas $1/p$ can be as large
as $e^N$. How does the \rhs of eq.~(40) manage to be exponentially large?
The crucial observation is that the mutual orthogonality condition~(39)
does not imply that $\j$ and $\Bar\c$ are normalized. Since the terms that
contribute to the chiral pole are of the form $\Bar\c\g_\m\j$, their magnitude
is not constrained by eq.~(39). In fact, on the lattice the off-diagonal terms
$\Bar\c\g_\m\j$ are exponentially large whenever the eigenvectors look like
bound states. As we have shown, for $p_\m\ll N^{-1}$ most of the
eigenvectors indeed look
like bound states. Thus, the pole on each domain wall receives contributions
from approximately half of the eigenstates.

  The physical picture that emerges from our discussion is the following.
In euclidean space, {\it the chiral
propagation on the domain wall is a coherent effect of the entire five
dimensional spectrum}. No single mode of the lattice
Dirac operator is capable of producing the chiral pole.

  Turning to the {\it interacting} theory we comment that,
roughly speaking, what is needed in order to have a chiral continuum limit
is that the fermionic determinant be
factorizable into four pieces. Two of them should depend on the
{\it four dimensional} gauge
fields in the vicinity of the (anti)-domain wall, while
the  other two should depend on the gauge degrees of freedom in the two
balk half-spaces. If this scenario works, we can expect that a continuum
limit should exists and that it should describes two chiral theories
(one on the domain wall
and one on the anti-domain wall) which no nothing of each other.
What might go wrong as a consequence of the
peculiarities of the euclidean spectrum is, for example, that non-local
correlations will develop in the $s$ direction. Another interesting possibility
is that a chiral continuum limit does exist, but that its {\it dynamics} is
qualitatively different from QCD.
Finding what really happens must await further investigations.

\vspace{5ex}
\centerline{\rule{5cm}{.3mm}}

\newpage
\centerline{\bf Acknowledgements}
\vspace{3ex}

  I thank A.~Casher for numerous helpful comments. I also benefited
from discussion with M.~Karliner and B.~Svetitsky. This research was
supported in part by the Basic Research Foundation administered by the Israel
Academy of Sciences and Humanities, and by a grant from the United States --
Israel Binational Science Foundation.

\vspace{5ex}
\centerline{\bf References}
\vspace{3ex}
\newcounter{00001}
\begin{list}
{[~\arabic{00001}~]}{\usecounter{00001}
\labelwidth=1cm}

\item J.~Smit, Nucl.~Phys.~B (Proc.~Suppl.) {\bf 17} (1990) 3.
M.F.L.~Golterman, Nucl.~Phys.~B (Proc.~Suppl.) {\bf 20} (1990) 515.
M.F.L.~Golterman, D.N.~Petcher and  J.~Smit, \NPB{370} (1992) 51.

\item L.H.~Karsten and J.~Smit, \NPB{183} (1981) 103.
H.B.~Nielsen and M.~Ninomiya, \NPB{185} (1981) 20.

\item D.B.~Kaplan, \PLB{288} (1992) 342.

\item J.~Goldstone and F.~Wilczek, \PRL{47} (1981) 986.
C.G.~Callan and J.A.~Harvey, \NPB{250} (1985) 427.

\item K.~Jansen, \PLB{288} (1992) 348.

\item K.~Jansen and M.~Schmaltz, {\it Critical Momenta of Lattice Chiral
Fermions}, UCSD/PTH 92-29.

\item M.F.L.~Golterman, K.~Jansen and D.B.~Kaplan, {\it Chern-Simons
Currents and Chiral Fermions on the Lattice}, submitted to \PRL.

\end{list}

\end{document}